\documentstyle[12pt,aasms4] {article}

\begin{document}

\title{Evidence for a Disk-Jet Interaction in the \\ Microquasar GRS 1915+105}

\author{Stephen S. Eikenberry\altaffilmark{1,3}, Keith Matthews\altaffilmark{1}, Edward H. Morgan\altaffilmark{2}, Ronald A. Remillard\altaffilmark{2}, Robert W. Nelson\altaffilmark{1}}

\altaffiltext{1}{California Institute of Technology, Pasadena, CA 91125}
\altaffiltext{2}{Massachusetts Institute of Technology, Cambridge, MA  02139}
\altaffiltext{3}{Sherman Fairchild Postdoctoral Fellow in Physics}

\begin{abstract}

		We report simultaneous X-ray and infrared (IR)
observations of the Galactic microquasar GRS1915+105 using XTE and the
Palomar 200-inch telescope on August 13-15, 1997 UTC.  During the last
two nights, the microquasar GRS 1915+105 exhibited quasi-regular
X-ray/infrared (IR) flares with a spacing of $\sim 30$ minutes.  While
the physical mechanism triggering the flares is currently unknown, the
one-to-one correspondence and consistent time offset between the X-ray
and IR flares establish a close link between the two.  At late times
in the flares the X-ray and IR bands appear to ``decouple'', with the
X-ray band showing large-amplitude fast oscillations while the IR
shows a much smoother, more symmetrical decline.  In at least three
cases, the IR flare has returned to near its minimum while the X-rays
continue in the elevated oscillatory state, ruling out thermal
reprocessing of the X-ray flux as the source of IR flare.
Furthermore, observations of similar IR and radio flares by Fender et
al. (1997) imply that the source of the IR flux in such flares is
synchrotron emission.  The common rise and subsequent decoupling of
the X-ray and IR flux and probable synchrotron origin of the IR
emission is consistent with a scenario wherein the IR flux originates
in a relativistic plasma which has been ejected from the inner
accretion disk.  In that case, these simultaneous X-ray/IR flares from
a black-hole/relativistic-jet system are the first clear observational
evidence linking of the time-dependent interaction of the jet and the
inner disk in decades of quasar and microquasar studies.

\end{abstract}


\section{Introduction}

	The galactic microquasar GRS1915+105 is one of the most
fascinating objects in astrophysics today.  Discovered as a transient
hard X-ray source (Castro-Tirado et al., 1992), GRS1915+105 is
best-known as the first Galactic source of superluminal jets (Mirabel
and Rodriguez, 1994).  The other known Galactic source of superluminal
jets, GRO1655-40, has been shown to harbor a compact object of mass
$\sim 7 M_{\odot}$ (Orosz and Bailyn, 1997), implying that both
GRO1655-40 and GRS1915+105 are powered by accretion onto a black hole.
The combination of relativistic jets and a central black hole has in
turn earned these two objects the name "microquasar", as they seem to
be the stellar-mass analogs of the massive black hole systems in
quasars and AGN.

	The microquasars offer an exciting opportunity to study the
physics of the black-hole/relativistic-jet interaction in ways that
are impossible in "normal" quasars.  In particular, the fact that
Galactic microquasars are much smaller than quasars means that they
can vary on much faster timescales -- fractions of a second to days
rather than weeks to years -- and the fact that they are closer by
orders of magnitude means that relatively fainter features can be
observed.  While Harmon et al. (1997) have shown a long-term
correlation between hard X-ray flux and jet activity in a microquasar,
the hard X-ray instruments lacked the sensitivity to study any fast
variability which might reveal the details of the interaction between
the jets and accretion.  Well-collimated outflows have also been
observed in a large variety of nonrelativistic systems, including
young stellar objects, planetary nebulae and accreting white dwarf
supersoft sources (see Livio 1997 for a recent review). All of these
systems have jet velocities comparable to the escape velocity from the
central object, which strongly suggests the jet must somehow tap
accretion energy being liberated in the very inner disk. However, the
actual mechanism which launches the jet is poorly understood, and no
detailed time-dependent correlation between the disk activity and jet
variability has ever been observed.  In the following sections, we
present X-ray and infrared (IR) observations of GRS1915+105 revealing
such fast variability which may have important implications for both
quasars and microquasars.  We then discuss the overall characteristics
of this variability, possible interpretations of this behavior, and
its ramifications for the black-hole/relativistic-jet interaction.  We
will give a more detailed analysis of the data and possible
theoretical interpretations in a forthcoming paper (Eikenberry et al.,
in preparation).  Finally we present the conclusions drawn from these
observations.

\section{Observations and Data Reduction}

	We observed GRS 1915+105 on the nights of 13-15 August 1997
UTC using the Palomar Observatory 200-inch telescope and the
Cassegrain near-infrared array camera in the K ($2.2 \mu$m) band.  We
configured the camera for high-speed operation, taking 64x64-pixel
(8x8-arcsec) images at a rate of 10 frames per second.  Absolute
timing was provided by a WWV-B receiver with $\sim 1$ ms accuracy.
The camera computer system limited us to 4000 consecutive frames
before restarting the integration.  We observed GRS 1915+105 in this
mode for approximately 5 hours each night, obtaining $\sim 1.5 \times
10^5$ frames per night.  The field of view in this mode was large
enough to capture both GRS 1915+105 and a nearby field star, Star A, which
has a magnitude of $K = 13.3$ mag (Eikenberry and Fazio, 1996; Fender
et al., 1997).  For each frame, we subtracted an averaged sky frame to
remove the array bias features which dominate the short exposures, and
then divided the result by a flat field.  For each star (GRS 1915+105 and
Star A), we then calculated the flux within a $\sim 1$-arcsec-radius
software aperture, and subtracted the sky flux from a surrounding
annulus.  We used the measured flux from Star A as a reference,
smoothing it with a 5-second boxcar filter to increase signal-to-noise
ratio, dividing it into the GRS 1915+105 flux, and then multiplying the
result by the 3.1 mJy flux density of Star A at $2.2 \mu$m.  We
present the resulting flux density for GRS 1915+105 on Aug. 14-15 UTC
with 1-second time-resolution in Figure 1.  We obtained X-ray
observations on the same nights using the PCA instrument on the Rossi
X-ray Timing Explorer (RXTE - see Greiner, Morgan, and Remillard
(1996) and references therein for further details regarding the
intrument and data modes).  Due to unfavorable Earth-Sun-GRS 1915
geometry at the time of the observations, only a limited fraction of
the IR observations had simultaneous X-ray coverage.  We present the
segments of data with significant ($>500$ sec) simultaneous coverage
in Figure 2.

\section{Discussion}

\subsection{Flaring Behavior}

	The most obvious features in Figures 1 and 2 are the large
amplitude X-ray/IR flares, which appear with a quasi-periodicity of
$\sim 30$ minutes on both August 14th and 15th.  Similar behavior has
been observed previously in the X-rays (Greiner, Morgan, and
Remillard, 1996), radio (Rodriguez and Mirabel, 1997; Pooley and
Fender, 1997), and in the IR (Fender et al., 1997).  However,
simultaneous coverage in both the X-rays and IR has never been
obtained until now.  The overlapping coverage here allows us to
establish that there is a one-to-one correspondence between the X-ray
and IR flares -- for every X-ray flare (of 7) there is a corresponding
IR flare, and vice versa.  While this could certainly be expected from
previous observations, this is the first conclusive evidence that the
X-ray and IR flares are caused by the same events.

	Except for the ubiquitous X-ray precursor spike, both the
X-ray and IR flare rises are relatively smooth and have essentially
monotonic rising edges.  In Figure 2, the IR flare peak is offset in
time relative to the X-ray flare peak.  Table 1 gives the time offsets
of the IR peak relative to the X-ray peak for the flares in Figure 2
that have sufficient simultaneous coverage.  In all 6 cases the time
offset between the peaks is consistent with a value of $\sim 300$
seconds, and a $1 / \sigma^2$-weighted combination of the measured
time offsets gives an average time offset between the peaks of $310
\pm 20$ seconds.  The time offset between the X-ray peak and the X-ray
precursor as well as the time offset between the beginning of the
X-ray rise and the X-ray peak are also constant within the
uncertainties given in Table 1.  Thus, the data do not support any
particular interpretation of the the value of the time offset between
the peaks (such as associating it with the light travel time between
the X-ray and IR-emitting regions).  Furthermore, it is interesting to
note that the X-ray precursor spikes appear to be coincident with the
beginning of the IR flares, suggesting that the spikes may be
associated with the initiation of the IR flares.  However, the
constancy of this offset and the one-to-one correspondence of the
flares are compelling evidence that the rising edges of the X-ray and
IR flares are closely linked, and possibly triggered by the same
event.

	Such a linkage is not present, however, in the latter stages
of the flares.  After reaching its peak flux, the X-ray flare enters a
wildly oscillating high state, with rapid large-amplitude variability
(changes of near 80\% of maximum flux on timescales of $\sim 10$
seconds).  These oscillations continue for several hundred to several
thousand seconds before the return to the X-ray low-state.  The IR
flares, on the other hand, appear to show decaying phases very similar
in their smoothness and timescale to the rising phases.  In
particular, no large amplitude oscillations are present in the IR.
Smaller amplitude ($\sim 10-30$\%) IR variability on $\sim 10$-second
timescales is present in some flares (e.g. the flare $\sim 9000$
sconds after MJD 50674.125 in Figure 1).  We will discuss these
``sub-flares'' in detail elsewhere, and only note here that this
variability is not seen in every flare (in contrast to the X-ray
oscillations), and when present it is not correlated with the X-ray
variability in any obvious way.

	We can also see that oscillations similar to those seen in the
X-rays are not simply ``smeared out'' in the IR by examining the
simultaneous observations near $t \sim 10000$ seconds and $t \sim
17500$ seconds after MJD 50674.125 and $t \sim 10000-12000$ seconds
after MJD 50675.125.  In all three cases, the X-ray oscillations
continue with a time-averaged flux near one-half of the flare maximum
while the IR flux has returned to near minimum.  To show this more
clearly, we took the X-ray and IR data from $t = 16000-18000$ seconds
after MJD 50674.125, subtracted the baseline from both bands, and
divided by the maximum flux level, resulting in the normalized flare
shown in Figure 3.  This figure clearly shows that the IR flux has
declined to near its minimum level while the X-rays remain in the
elevated oscillatory state.  Thus, we conclude that the X-ray and IR
emissions do indeed ``decouple'' from one another during the decay
phases of the flares.

\subsection{Interpretation of the Flares}

	This behavior is inconsistent with a scenario wherein the
majority of the IR flux arises from thermal reprocessing of a fraction
of the X-ray flux on structures near the black hole (the disk and/or
companion star).  In such a scenario, radiative delays are negligible,
but light travel time effects cause both delay and smearing of the
reprocessed flux as compared to the X-ray, as is seen in the optical
counterparts of X-ray bursts from X-ray binaries (Pedersen et al.,
1982).  This is in direct opposition to the observed behavior in
Figure 3, where the IR flare continues to drop while the X-rays
oscillate around a high average value.  Simultaneous X-ray/IR
observations on Aug. 13 1997 UT also contradict the thermal
reprocessing scenario, having high X-ray count rates of $\sim 3 \times
10^4 \ {\rm s^{-1}}$ but IR flux densities of $\sim 5$ mJy -- only
slightly above the minimum seen here.  Thus, we conclude that thermal
reprocessing of the X-ray flux cannot be the primary source of IR flux
during the flares.

	On the other hand, the observed behavior of the X-ray/IR
flares is consistent with a scenario wherein the IR flux arises from
ejected synchrotron-emitting plasma.  As noted above, the one-to-one
correspondence between the X-ray/IR flares and the constant time
offset between the X-ray/IR peaks indicate that the same event
triggers both flares, and thus imply that the IR- and X-ray-emitting
regions are initially physically close to each other.  The subsequent
decoupling of the IR emission from the X-rays implies that a
significant separation between the IR-emitting and X-ray-emitting
regions develops at later times.  Since the post-flare X-ray
oscillations continue to show peak luminosities of $\sim 10^{39}$
ergs/s, the X-ray emission region must remain very close to the
compact object, and is very likely the inner disk.  Therefore, we
conclude that the IR-emitting plasma arises from the inner disk and is
ejected from the system.

	This scenario is also supported by same-day IR and radio
observations of previous IR flares from GRS 1915+105 (Fender et al.,
1997).  The high brightness temperature of the radio flares ($\sim
10^{10}$ K) rules out a thermal origin.  This fact, together with the
flat radio spectrum of the flares (pooley and Fender, 1997), and the
linear polarization of the source during flaring activity ($\sim 1
\%$; Rodriguez and Mirabel, 1997) strongly implies a synchrotron
origin for the flares.  The decay timescales are similar for both the
radio and IR flares, implying that adiabatic expansion may be the
dominant cooling mechanism.  Thus, these observations also favor the
interpretation of the IR flares as emission from ejected plasma.
Further radio observations (Pooley and Fender, 1997) revealed
wavelength-dependent delays in the peaks of the radio flares.  If the
IR-radio flares are due to expanding ejecta, it is tempting to explain
such a delay as the transition from optically thick to optically thin
synchrotron emission at longer and longer wavelengths.  Such an
interpretation might also explain why the X-ray/IR time offset between
the peaks here is 300 seconds, while at other times the X-ray/radio
peak separation appears to be closer to $\sim 800$ seconds (Pooley and
Fender, 1997).  However, while this explanation agrees qualitatively
with the observations, it is not clear that the observed time offset
scaling versus frequency is compatible with such a transition in an
adiabatically-expanding plasma.

	The plasma ejection scenario is by no means the only possible
explanation for the behavior shown above.  The salient features are
that the X-ray and IR flares are triggered by the same event, but are
clearly decoupled from one another at late times in the flare.
However, given that GRS 1915+105 is {\it known} to eject
synchrotron-emitting blobs of relativistic plasma (Mirabel and
Rodriguez, 1994), the ejection hypothesis seems the most likely and
natural explanation for the behavior observed here.  We note, however,
that even if the flares are due to plasma ejection, they are NOT
identical to the superluminal events observed from GRS 1915+105.  The
timescales of the flares, on the order of 2000 seconds (see Figure 1),
is significantly faster than the several-day timescale seen during
superluminal ejections (Mirabel and Rodriguez, 1994).  Furthermore,
the radio fluxes of flares similar to those in Figure 1 are lower than
the major ejections by an order of magnitude (Fender et al., 1997).
Thus, any plasma ejection taking place in the 14-15 August 1997
observations can at most be a ``baby jet'' analog to the much larger
superluminal ejection events.

\section{Conclusion}

	We have seen simultaneous X-ray/IR flares from GRS 1915+105
with a quasi-regular spacing of $\sim 30$ minutes.  The one-to-one
correspondence of the X-ray flares with the IR flares, and the
constant time offset between them establishes that both flares
originate from the same event.  At late times in the flare, the X-ray
and IR flares appear to decouple, ruling out thermal reprocessing of
the X-rays as the source of the IR flares.  This behavior is
consistent with ejection of an IR-emitting plasma from the system
during increased activity in the inner accretion disk.

	If the simultaneous X-ray/IR flares are indeed the signatures
of plasma ejection in GRS 1915+105, then they are giving us insights
into the ``central engine'' of a black-hole/relativistic-jet system.
Despite the fact that quasar and AGN jets have been studied for
decades, no observations of the central jet engine have been possible
due to the great distances to the quasars.  Since microquasars are
much smaller, closer, and vary faster than the extragalactic systems,
they have been considered as potential ``laboratories'' for the study
of the physics of black-hole/relativistic-jet systems.  From the
observations reported here, it appears that the microquasars have
lived up to this potential.  In particular, these observations
intimately link the apparent plasma ejection to activity in the inner
accretion disk -- an important constraint for the multitude of
theoretical models attempting to describe black-hole/relativistic-jet
systems.  More detailed analyses of this rich data set and further
multi-wavelength observations of GRS 1915+105 will undoubtedly yield
even more insight into the physical characteristics and behaviors of
the central engines.

\acknowledgements The authors wish to thank G. Neugebauer for useful
discussions of these observations.  SE acknowledges the support of a
Sherman Fairchild Postdoctoral Fellowship in Physics.

\begin{deluxetable} {lccc}
\tablecolumns{4} 
\tablewidth{0pc} 
\tablecaption{Time offsets between the X-ray peaks and IR
peaks for the flares shown in Figure 2.  The X-ray and IR time series
were rebinned to 10-second resolution (to smooth over the rapid
small-scale variability), and peaks determined as the time when the
flare reaches maximum flux.  The uncertainty in this value is
determined by the time span over which the flux is equal to the
maximum within the statistical uncertainties (see Figures 1 and 2),
and thus depends on both the sharpness of the flare peak and the
statistical uncertainties in the flux measurements.}  
\tablehead{
\colhead{MJD} & \colhead{Flare Time\tablenotemark{a} \  (s)} &
\colhead{Offset (s)} & \colhead{Unc. (s)}} 
\startdata 
50674 & 4000 & 400 & $^{+50}_{-110}$ \nl 
50674 & 11000 & 345 & $\pm 30$ \nl 
50674 & 16000 & 285 & $\pm 30$ \nl 
50675 & 4000 & 220 & $\pm 160$ \nl 
50675 & 15500 & 305 & $^{+20}_{-70}$ \nl 
50675 & 17000 & 255 & $\pm 160$ \nl
\enddata

\tablenotetext{a}{Approximate start of the X-ray flare as measured from
3 UTC on given date (= MJD + 0.125).}

\end{deluxetable}

\begin{figure}
\vspace*{180mm}
\includegraphics{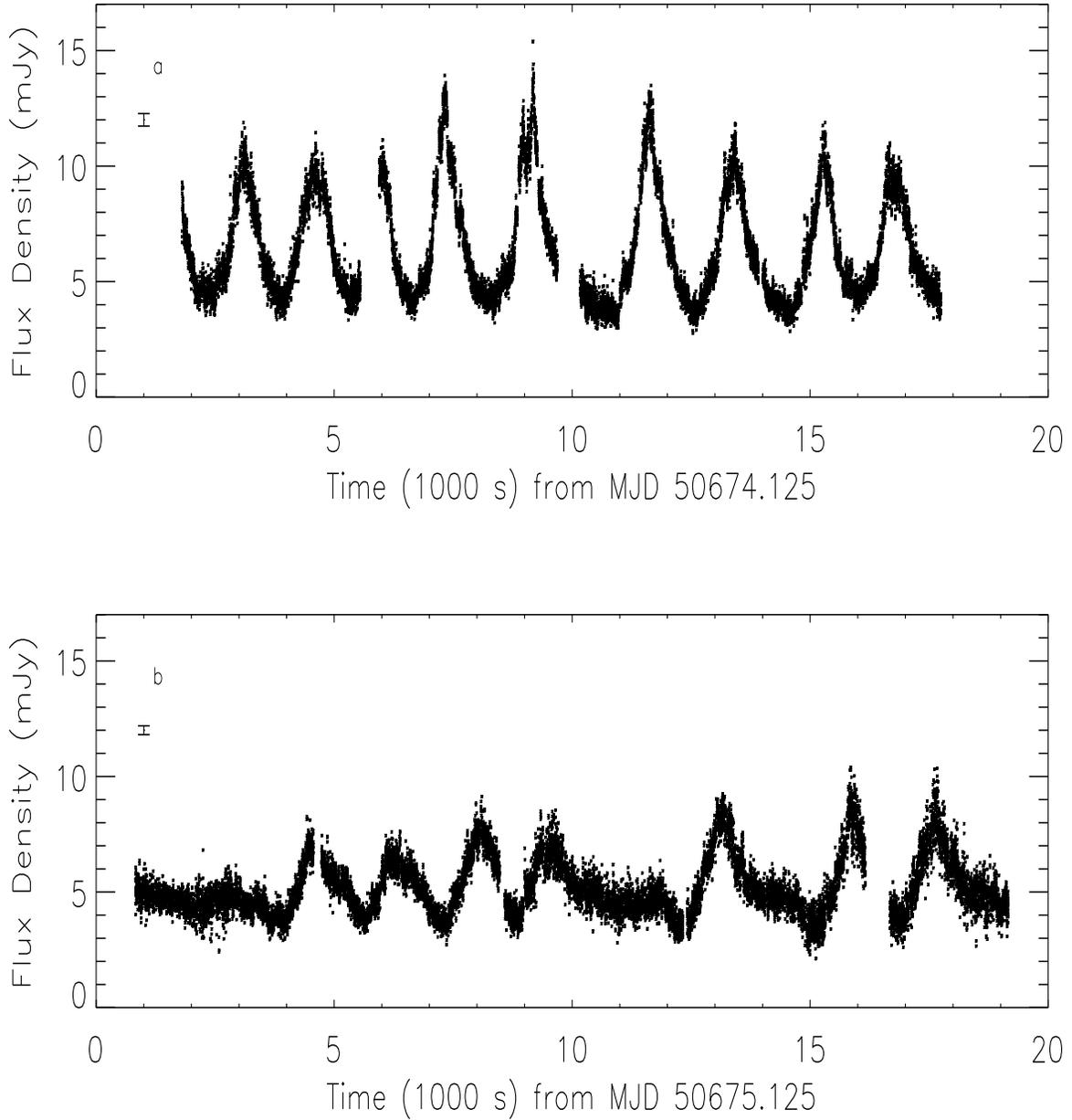}
\caption{Infrared flux density from GRS 1915+105 measured with 1-second time-resolution in the K ($2.2 \mu$m) band from the Palomar 200-inch telescope.  Typical uncertainties ($\pm 1 \sigma$) are shown - they are dominated by the uncertainty in the reference flux density from Star A, and are calculated by taking the standard deviation in that flux density and multiplying by the ratio of the GRS 1915+105 flux density to the reference flux density.}
\end{figure}

\begin{figure}
\vspace*{180mm}
\includegraphics{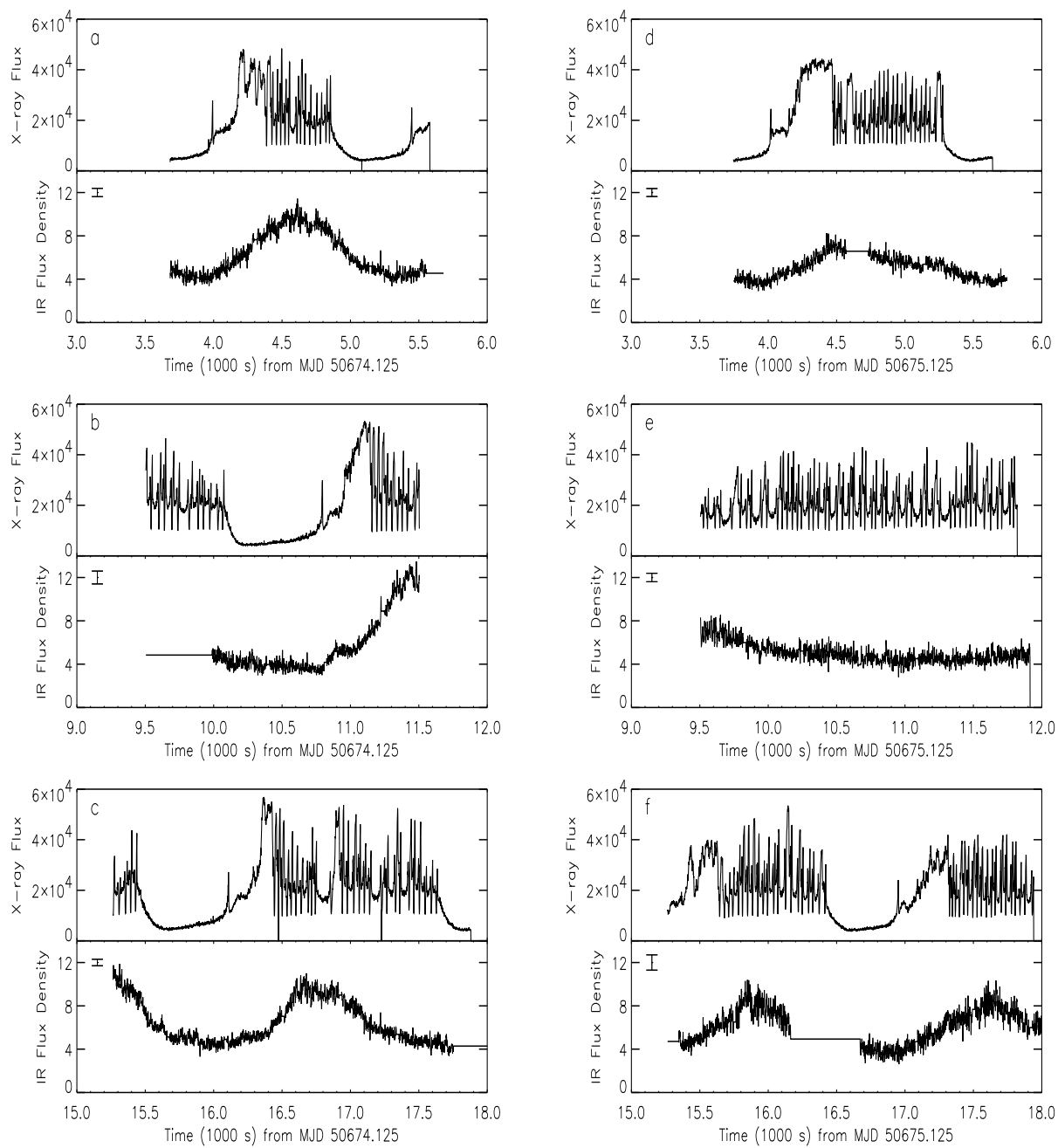}
\caption{Simultaneous X-ray and infrared observations of GRS 1915+105 at 1-second time-resolution.  X-ray flux is in PCA counts/s, IR flux density is in milliJansky.  Typical uncertainties are shown for the IR.  The X-ray uncertainties are assumed to be Poissonian, and are too small to be seen on this scale.}
\end{figure}

\begin{figure}
\vspace*{180mm}
\includegraphics{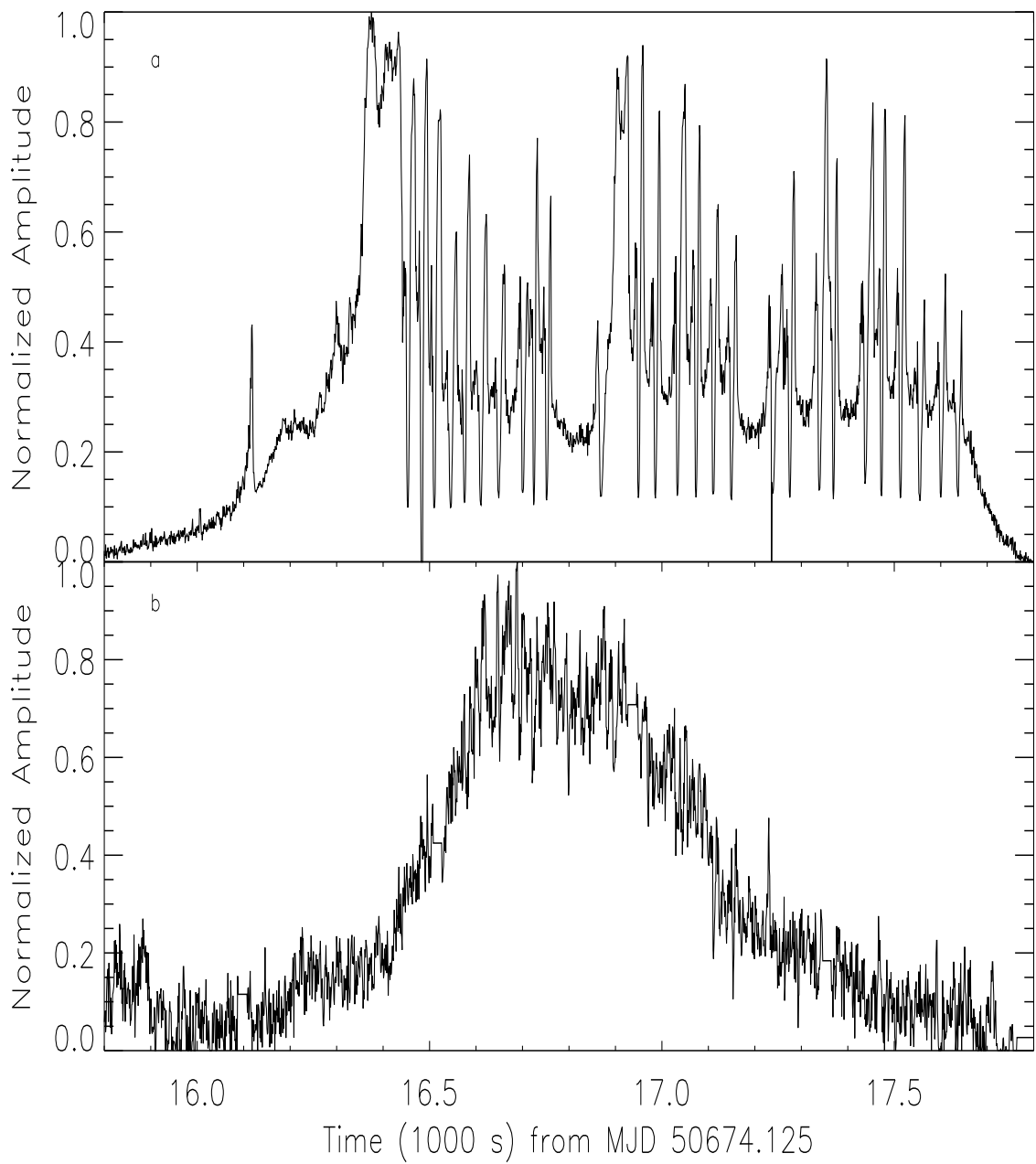}
\caption{Simultaneous observations of a flare in the (a) X-ray, (b) IR.  Both bands have a constant baseline subtracted and are normalized to a maximum amplitude of 1.0.  Note that the IR flare has returned to near its baseline level, while the X-rays continue to oscillate wildly with a time-averaged flux near one-half of the flare maximum.  This behavior rules out thermal reprocessing of the X-rays as the primary source of IR flux in the flare.}
\end{figure}

\end{document}